\documentclass[preprint,nofootinbib]{revtex4}
\usepackage{amssymb}
\usepackage{amsfonts}
\usepackage{amsmath}
\usepackage{graphicx}
\usepackage[usenames,dvipsnames,svgnames,table]{xcolor}

\usepackage{hyperref}
\hypersetup{
colorlinks,breaklinks,
            urlcolor=[rgb]{0.8,0.2,0.5},
            citecolor=[rgb]{0.5,0,0},
}

\makeatletter \@addtoreset{equation}{section}

\makeatother
\def\be{\begin{equation}}   \def\ee{\end{equation}}

\begin{document}
\title{Holography of 3D asymptotically flat  black holes}
\author{Reza Fareghbal$^{a,b}$}
\email{r$_$fareghbal@sbu.ac.ir}
\author{Seyed Morteza Hosseini$^{a,c,d}$}
\email{morteza.hosseini@mib.infn.it}
\affiliation{$^{a}$Department of Physics, Shahid Beheshti University G.C., Evin, Tehran 19839, Iran}
\affiliation{$^{b}$School of Particles and Accelerators, Institute for Research in Fundamental Sciences (IPM), P.O. Box 19395-5531, Tehran, Iran}
\affiliation{$^{c}$Dipartimento di Fisica, Universit\`a di Milano-Bicocca, I-20126 Milano, Italy}
\affiliation{$^{d}$INFN, sezione di Milano-Bicocca, I-20126 Milano, Italy}

\begin{abstract}
We study the asymptotically flat rotating hairy black hole solution
of a three-dimensional gravity theory which is given
by taking the flat-space limit (zero cosmological constant limit) of new massive gravity.
We propose that the dual field theory of the flat-space limit of new massive gravity can
be described by a contracted conformal field theory which is
invariant under the action of the BMS$_{3}$ group.
Using the flat/contracted conformal field theory correspondence, we construct a stress tensor
which yields the conserved charges of the asymptotically flat black hole solution.
We check that our expressions of the mass and angular momentum
fit with the first law of black hole thermodynamics.
Furthermore, by taking the appropriate limit of the Cardy formula in the parent conformal field theory,
we find a Cardy-like formula which reproduces the Wald\rq{}s entropy
of the 3D asymptotically flat black hole.
\end{abstract}
\maketitle

\hypersetup{
colorlinks,breaklinks,
            linkcolor=black
}

\tableofcontents

\hypersetup{
colorlinks,breaklinks,
            linkcolor=[rgb]{0.1,0.4,0.1}
}

\section{Introduction}
Taking the flat-space limit (zero cosmological constant limit)
of asymptotically AdS spacetimes results in asymptotically flat geometries.
This procedure can be done by taking
the $\ell\to\infty$ limit where $\ell$ is the radius of AdS spacetime.
From the field theory perspective,
one could expect that the $\ell\to\infty$ limit in the bulk theory
has a holographic description at the boundary.
Recently, it has been argued that the flat-space limit of AdS gravity
is dual to the $\dot {\rm I}$n$\ddot {\rm o}$n$\ddot {\rm u}$-Wigner
contraction of the boundary conformal field theory (CFT) \cite{Bagchi:2010zz,Bagchi:2012cy}.\footnote{It is worth noticing
that another interesting approach to understanding
flat space quantum gravity is given in \cite{Krishnan:2013wta}.
Therein, the authors showed that interpreting the inverse AdS$_{3}$ radius $1/\ell$
as a Grassmann variable leads to a map from gravity in AdS$_{3}$ to gravity in flat space.}

The so-called flat/contracted conformal field theory (CCFT) correspondence has received a great deal of attention recently.
For example, in \cite{Bagchi:2012xr} a Cardy-like formula has been obtained for the
two-dimensional CCFT which yields the correct entropy of the three-dimensional cosmological solution.
These asymptotically flat spacetimes can be obtained by taking the flat-space limit,
as in \cite{Cornalba:2002fi}, of nonextremal Ba\~{n}ados-Teitelboim-Zanellii (BTZ) black holes.
After taking the flat-space limit, the outer horizon of BTZ is mapped to infinity;
however, the value of the inner horizon remains finite and defines the cosmological horizon. 
The entropy of the cosmological solution has been identified with the area of the cosmological horizon.
In the literature (see for example \cite{Detournay:2012ug}), a modified Cardy formula has been introduced
which reproduces the entropy of the inner horizon of BTZ black holes.
The CFT origin of this formula has not been well understood yet, but the observation of \cite{Fareghbal:2014qga,Riegler:2014bia}
is that if we accept the modified Cardy formula related to the inner horizon of the BTZ
and contract it by using appropriate parameters of CCFT, the final result
is exactly the CCFT Cardy-like formula which yields the entropy of the cosmological horizon.

Furthermore, in \cite{Fareghbal:2013ifa}, the authors found the correlation functions
of CCFT energy-momentum tensor by using the contraction of CFT correlation functions
for finding the quasilocal stress tensor of the asymptotically
flat spacetimes which gives the correct conserved charges of these geometries.

The Flat/CCFT correspondence can also propose a dual field theory
which lives at the horizon of nonextreme black holes.
The idea begins from the appearance of Rindler spacetime in the near horizon limit of
nonextreme black holes.
If one starts with the Rindler-AdS/CFT correspondence \cite{Czech:2012be,Parikh:2012kg} and takes
the flat-space limit in the bulk, which results in the Rindler spacetime,
the boundary field theory is given by the contraction of the parent CFT.
This proposal has been used in \cite{Fareghbal:2014oba}.
Therein, the authors found the Bekenstein-Hawking entropy of the nonextreme BTZ black hole
by counting the CCFT microsates.

Moreover,   Bagchi {\it et al.}\,
calculated the entanglement entropy of a two-dimensional field theory with
Galilean conformal symmetry\footnote{The group of symmetries of CCFT$_{2}$ is isomorphic to Galilean Conformal Algebra
but in higher dimensions these are not the same.} recently \cite{Bagchi:2014iea}.
The authors used the Wilson lines approach and found the holographic entanglement entropy
they computed is in precise agreement with the ones obtained in the field theory side. 
For an almost complete list of papers related to the Flat/CCFT correspondence, see the references of \cite{Bagchi:2014iea}.

In (2+1)-dimensional Einstein gravity, black holes can exist only in the
presence of a negative cosmological constant \cite{Ida:2000jh}.
In order to find asymptotically flat black holes in three dimensions one has  to consider higher-derivative gravity theories.
The entropy of these black holes can be obtained by Wald\rq{}s formula \cite{Wald:1993nt}.
A successful theory of quantum gravity should be able to give a microscopic description of this semiclassical entropy.
An alternative approach for this study is using holography.
The Falt/CCFT correspondence as a duality between quantum gravity in the asymptotically flat backgrounds
and a field theory with contracted conformal symmetry can be an appropriate context to study this problem.
The current paper is focused in this direction.

We consider a theory of gravity which is given by taking the flat-space limit of  new massive gravity (NMG) \cite{Bergshoeff:2009hq}.
This theory possesses remarkable properties.
In \cite{Deser:2009hb}, it was shown that it is a ghost-free and power-counting UV finite, three-dimensional gravity.
We use the dictionary of the Flat/CCFT correspondence for finding quasi local stress tensor
of the new type of asymptotically flat black hole \cite{Oliva:2009ip}.
Using the holographic stress tensor along with the Brown and York\rq{}s method \cite{Brown:1992br},
we compute the conserved charges of this black hole.
We also take the limit from the Cardy formula and find a Cardy-like formula for the
CCFT and show that this gives agreement with the Wald\rq{}s semiclassical approach.

The next sections are devoted to two main parts.
In Sec.(\ref{Bulk Solutions}) we introduce the bulk solutions.
We start from NMG and review its  asymptotically AdS rotating hairy black hole.
Then we take the flat-space limit from the
action and its black hole solution and introduce the asymptotically
flat rotating hairy black hole with some novel properties.
We calculate its entropy using Wald\rq{}s formula and verify for it the first law
of black hole thermodynamics.
In Sec.(\ref{Dual Boundary Theory}) we argue about the dual boundary theory of the bulk solution.
We shortly review the known results about the dual CFT of NMG
and then try to contract these results and find a CCFT which is dual to the
higher-derivative gravity theory of \cite{Deser:2009hb}.
This work is another check for the correctness of the Flat/CCFT correspondence.

\section{Bulk Solutions}
\label{Bulk Solutions}

\subsection{Rotating hairy black hole of NMG}
We consider the three-dimensional higher-derivative gravity theory known as NMG.
This theory is described by the parity-invariant action \cite{Bergshoeff:2009hq}
\be\label{BHTaction}
S=\frac{1}{16\pi G}\int d^{3}x \sqrt{-g}\left[R-2\Lambda+\frac{1}{m^{2}}K\right],
\ee
where
\be
K=R_{\mu\nu}R^{\mu\nu}-\frac{3}{8}R^{2}.
\ee

The above theory (\ref{BHTaction}) for the special case $m^{2}\ell^{2}=1/2$
has the following rotating black hole solution \cite{Oliva:2009ip}
\be\label{BHTbh}
ds^{2}=-NFdt^{2}+\frac{dr^{2}}{F}+r^{2}\left(d\phi+N^{\phi}dt\right)^{2},
\ee
where $N$, $N^{\phi}$, and $F$ are  given by
\begin{align}
N  &  =\left[  1+\frac{b\ell^{2}}{4H}\left(  1-\Xi^{\frac{1}{2}}\right)  \right]
^{2}\ ,\nonumber\\
N^{\phi}  &  =-\frac{a}{2r^{2}}\left(  4GM-bH\right)  \ ,\label{Ns&F}\\
F  &  =\frac{H^{2}}{r^{2}}\left[  \frac{H^{2}}{\ell^{2}}+\frac{b}{2}\left(
1+\Xi^{\frac{1}{2}}\right)  H+\frac{b^{2}\ell^{2}}{16}\left(  1-\Xi^{\frac{1}{2}
}\right)  ^{2}-4GM\ \Xi^{\frac{1}{2}}\right]  \ ,\nonumber
\end{align}
and
\begin{align}
\nonumber H&=\left[  r^{2}-2GM\ell^{2}\left(  1-\Xi^{\frac{1}{2}}\right)  -\frac{b^{2}\ell^{4}
}{16}\left(  1-\Xi^{\frac{1}{2}}\right)  ^{2}\right]  ^{\frac{1}{2}},
\label{H}\\
\Xi&=1-a^{2}/\ell^{2}.
\end{align}
It is labeled by three parameters:
the mass $M$, the angular momentum $J=M a$, and an additional \lq\lq{}gravitational hair\rq\rq{} parameter $b$.
The rotation parameter $a$ is bounded according to $-\ell\leq a\leq \ell$.

The angular velocity of the horizon is
\be
\Omega_{+}=\frac{1}{a}\left(  \Xi^{\frac{1}{2}}-1\right)  \ . \label{Omega+}
\ee
We can associate a Hawking temperature and entropy to it
\begin{align}
T&=\frac{\Xi^{\frac{1}{2}} }{\pi \ell}\sqrt{2G\Delta M \left( 1+\Xi^{\frac{1}{2}} \right) ^{-1}},  \label{Temperature}\\
S&=\pi \ell\sqrt{\frac{2}{G}\Delta M
\left( 1+\Xi^{\frac{1}{2}} \right) },  \label{Entropy}
\end{align}
where
\be
\Delta M=M+\frac{b^{2}\ell^{2}}{16G}\ .
\ee
These quantities fulfill the relation
\be
TdS=\Xi^{\frac{1}{2}} \text{ }dM+\frac{b\ell^{2}}{8G}\Xi^{\frac{1}{2}} 
\text{ }db-\frac{1}{a}\left( 1-\Xi^{\frac{1}{2}} \right) \Delta M \text{ }da\ .  \label{calentate}
\ee
Not much more work would bring this equation to the familiar form of the first law of black hole thermodynamics
\be
d(\Delta M)=TdS-\Omega_{+}d(\Delta J)\, ,
\ee
where we defined
\be
\Delta J=a \Delta M\, .
\ee

\subsection{The flat-space limit of NMG and its black hole solution}
In order to have a well-defined flat-space limit ($\Lambda\to 0$ or $\ell\to\infty$)
for \eqref{BHTaction} in the special point $m^2\ell^2=1/2$, we also need to scale Newton\rq{}s constant to infinity while keeping fixed $\kappa=\ell^{2}/G$ .
Thus, the flat-space limit of NMG action \eqref{BHTaction} becomes
\be\label{Flataction}
S=\frac{\kappa}{8\pi}\int d^{3}x \sqrt{-g} K.
\ee
Moreover, a well-defined flat-space limit for the black hole solution \eqref{BHTbh}
needs a scaling of the mass parameter $M$ such that $\mu=M G$ remains fixed.
The final line element for the asymptotically flat rotating hairy black hole is given by 
 \be\label{FlatBH}
ds^{2}=-{\cal F}dt^{2}+\frac{r^{2}}{{\cal F}\Delta}dr^{2}+a{\cal F}dtd\phi+r^{2}d\phi^{2},
\ee
where $\Delta(r)$ and ${\cal F}(r)$  are functions of the radial coordinate $r$, given by
\begin{align}
\Delta &  = r^{2}- \mu a^{2}-\left(\frac{a^{2}b}{8}\right)^{2} \ ,\nonumber\\
{\cal F} &  = b\sqrt{\Delta}-4 \mu\,.
\end{align}
The Ricci scalar of this asymptotically flat black hole can be written as
\be
R=-\frac{16\, b}{a^2 b+8\sqrt{\Delta}}\, .
\ee
One can verify that \eqref{FlatBH} satisfies the
equations of motion resulting from the action \eqref{Flataction}.
It is worth noting that in \cite{Deser:2009hb}, it was argued that the three-dimensional gravity
theory described by \eqref{Flataction} is  ghost free and finite.

Horizons of \eqref{FlatBH} are at
\be
r_{+}= \frac{a^2 b}{8}+\frac{4\mu}{b}\, ,
\qquad r_{-}=\sqrt{\left(\frac{a^{2} b}{8}\right)^{2}+\mu a^{2}}\, ,
\ee
and one can calculate the entropy of the outer horizon using the Wald\rq{}s formula.
This formula gives the black hole entropy in an arbitrary diffeomorphism invariant theory and is given by
\be\label{Wald}
S=-\frac{2\pi}{16\pi G}\int_{\Sigma_{h}}\,\frac{\delta L}{\delta R_{\alpha\beta\gamma\delta}}\,
\epsilon_{\alpha\beta}\,\epsilon_{\gamma\delta}\,\bar{\epsilon}\ ,
\ee
where $L$ is the Lagrangian, and $\bar{\epsilon}$, $\epsilon_{\mu\nu}$,
denote the volume form and the binormal vector to the spacelike
bifurcation surface $\Sigma_{h}$, respectively.
$\epsilon_{\mu\nu}$ is normalized as $\epsilon^{\mu\nu}\epsilon_{\mu\nu}=-2$.
For the action \eqref{Flataction} and the asymptotically flat rotating hairy black hole solution \eqref{FlatBH} we obtain 
\begin{subequations}
\begin{align}
\frac{\partial L}{\partial R_{\alpha\beta\gamma\delta}}
&=\frac{3}{8}R
\left(g^{\alpha\delta}g^{\beta\gamma}-g^{\alpha\gamma}g^{\beta\delta}\right)
+\frac{1}{2}\left(g^{\alpha\gamma}R^{\beta\delta}
-g^{\alpha\delta}R^{\beta\gamma}
-g^{\beta\gamma}R^{\alpha\delta}
+g^{\beta\delta}R^{\alpha\gamma}\right),\\
\epsilon_{\alpha\beta}&=-
\left(\frac{a^{2}{\cal F}+4r^{2}}{\Delta}\right)^{\frac{1}{2}}
\delta^{t}_{[\alpha}\delta^{r}_{\beta]}\, .
\end{align}
\end{subequations}
Therefore, the Wald\rq{}s entropy for the new type of asymptotically flat black hole becomes
\be\label{flat entropy}
S_{{\rm flat}}=\frac{\pi\kappa b}{2}\, .
\ee

It is instructive to derive the above entropy by taking the flat-space limit of the entropy \eqref{Entropy}.
The entropy can, therefore, be computed as follows:
\begin{align}\label{FEntropy}
\lim_{\ell\to\infty} S&=\lim_{\ell\to\infty}\pi \ell\sqrt{\frac{2}{G}\Delta M\left(  1+\Xi^{\frac{1}{2}}\right)  }\nonumber\\
&=\lim_{\ell\to\infty}\frac{\pi\kappa b}{2}\sqrt{1+\frac{16\mu}{b^{2}\ell^{2}}}\nonumber\\
&=\frac{\pi\kappa b}{2}=S_{{\rm flat}}\, .
\end{align}
Now, consider the Hawking temperature.
From (\ref{Temperature}) it follows that
\begin{align}\label{FTemperature}
\lim_{\ell\to\infty}T  &  =\lim_{\ell\to\infty}\frac{1}{\pi\ell}\Xi^{\frac{1}{2}}\sqrt{2G\Delta M\left(  1+\Xi^{\frac
{1}{2}}\right)  ^{-1}}\ ,\nonumber\\
&=\lim_{\ell\to\infty}\frac{b}{4\pi}\sqrt{1+\frac{16\mu}{b^{2}\ell^{2}}}\nonumber\\
&=\frac{b}{4\pi}=T_{{\rm flat}}\, .
\end{align}
These quantities fulfill the relation
\be\label{Fcalentate}
T_{\text{flat}}\,dS_{\text{flat}}=\frac{b \kappa}{8}\,db\, .
\ee
This agrees precisely with the $\ell\to\infty$ limit of (\ref{calentate}).
A direct calculation by using \eqref{FlatBH} or taking the flat-space limit of \eqref{Omega+}
shows that the angular velocity of the black hole \eqref{FlatBH} at the outer horizon
is zero ($\Omega_{{+}_{\rm flat}}=0$) though it has a nonvanishing angular momentum.

From \eqref{flat entropy} and \eqref{FTemperature}, it is clear that the hair parameter $b$
determines the entropy and the temperature of the outer horizon.
In the $b\to 0$ limit the hairy black hole \eqref{FlatBH} is reduced to the cosmological solution of \cite{Cornalba:2002fi}.
In this limit, $r_{+}$ is mapped to infinity; however, $r_{-}$ remains finite and defines the cosmological horizon. 

\section{Dual Boundary Theory}
\label{Dual Boundary Theory}

\subsection{CFT dual to NMG}
In \cite{Bergshoeff:2009aq, Giribet:2009qz}, it was proposed that NMG has a dual description in terms of a CFT.\footnote{In 2013, de Buyl {\it et al.}
considered the asymptotically dS case, $\Lambda=+1/\ell^{2}>0$,
within the context of dS/CFT correspondence \cite{deBuyl:2013ega}.
We would like to thank the referee for bringing this paper to our attention.}
The charges associated to the asymptotic symmetries enhance the isometry of asymptotically $AdS_3$ spacetimes
to two copies of the Virasoro algebra.
The central charges are given by
\be
c_{\pm}=c=\frac{3\ell}{2G}\left(1+\frac{1}{2m^{2}\ell^{2}}\right)\ .\label{central charge}
\ee
At the spacial point  $m^{2}\ell^{2}=1/2$, the central charges are twice
the values proposed by Brown and  Henneaux for the Einstein gravity
with negative cosmological constant \cite{Brown:1986nw}, i.e.,
\be
c=\frac{3\ell}{G}\ .\label{central charge}
\ee
The entropy of the black hole \eqref{BHTbh} can be given by the Cardy formula
\be
S=2\pi\sqrt{\frac{c_{+}\Delta_{+}}{6}}+2\pi\sqrt{\frac{c_{-}\Delta_{-}}{6}
}\ ,\label{Cardy formula}
\ee
where $\Delta_\pm$ are the eigenvalues of the left and right Virasoro generators $L_{0}^{\pm}$ and are given by
\be
\Delta_{\pm}=\frac{1}
{2}\Delta M\left(  \ell\pm a\right)  \ .\label{Delta+-}
\ee
Using (\ref{central charge}) and (\ref{Delta+-}), this is
\be
S =\pi \ell\sqrt{\frac{2}{G}\left(  1+\Xi^{\frac{1}{2}}\right)  \Delta M}\ ,
\ee
in precise agreement with  (\ref{Entropy}).

\subsection{CCFT dual to the flat-space limit of NMG}
In this section we want to propose a dual description for the theory of gravity given by \eqref{Flataction}.
To do so, we will use the idea which was first proposed in papers \cite{Bagchi:2010zz,Bagchi:2012cy}.
That is, if we start from the AdS/CFT correspondence,
the large AdS radius limit in the bulk is equivalent to a contraction
of spacetime coordinates in the boundary CFT.\footnote{We refer to \cite{Bagchi:2012cy} for
a full-scale investigation into the CCFT representation.}

We shall first show how one obtains the appropriate coordinate which must be contracted in the parent CFT.
Let us look at the conformal boundary of the black hole \eqref{BHTbh} for an arbitrary large $\ell$.
It could be written as follows:
\be\label{CB}
ds_{{\rm C.B.}}^{2}=\frac{r^{2}}{\kappa^{2}}\left(-\frac{\kappa^{2}}{\ell^{2}}dt^{2}+\kappa^{2}d\phi^{2}\right)\, .
\ee
We have  used $\kappa$ in the conformal factor to make it dimensionless.
Moreover, the fact that $\kappa$ is fixed in our flat-space limit makes
the conformal factor well defined for all large values of $\ell$.
Now, $\ell$ can be absorbed by defining new time coordinate as $\tau=\kappa t/\ell$.
The dual CFT lives on a cylinder with coordinates $(\tau, \phi)$ and radius $\kappa$.
Taking the $\ell\to\infty$ limit (or $\kappa/\ell\to 0$ limit),
it is obvious that the flat-space limit in the bulk induces a contraction in
$t$ of the boundary CFT reducing it to the two-dimensional CCFT.

\subsubsection{Symmetries of CCFT}
According to the proposal of \cite{Bagchi:2010zz,Bagchi:2012cy}, the symmetries of CCFT
realize the group of  asymptotic symmetries of the asymptotically flat  spacetimes at null infinity,
namely the Bondi-Metzner-Sachs (BMS) group \cite{BMS,Barnich aspects}.
There is a very precise procedure, called the $\dot {\rm I}$n$\ddot {\rm o}$n$\ddot {\rm u}$-Wigner contraction, by which one
can obtain the CCFT algebra from the relativistic conformal algebra of the parent CFT.
Let us consider two copies of the Virasoro algebra\footnote{The relativistic conformal algebra
consists of two copies of the Virasoro algebra.}
\begin{align}\label{Virasoro}
  \nonumber [L^+_m,L^+_n]&=(m-n)L^+_{n+m}+{c^+\over 12}m(m^2-1)\delta_{m+n,0}\, ,\\
  \nonumber [L^-_m,L^-_n]&=(m-n)L^-_{n+m}+{c^-\over 12}m(m^2-1)\delta_{m+n,0}\, ,\\ 
            [L^+_m,L^-_n]&=0\, .
  \end{align}  
For a small parameter $\epsilon$, at the level of the algebra, if we define\footnote{It is clear from (3.8)
that other positive powers of $\epsilon$ are meaningless
since one can redefine $\epsilon$ and rewrite them finally as in (3.8).
This definition is consistent with the observation of \cite{bar-comp}.
That is, one can obtain the BMS$_{3}$ algebra by taking the flat-space limit
from the asymptotic symmetry algebra of three-dimensional asymptotically AdS spacetimes.}
\begin{align}\label{def of BMS generators}
\nonumber L_n &=L^+_n-L^-_{(-n)}\, ,\\
          M_n &=\epsilon\left(L^+_n+L^-_{(-n)}\right), 
\end{align}
we can see that the CCFT algebra is generated from the Virasoro algebras, on taking the $\epsilon\to0$ limit, i.e.,
\begin{align}\label{BMS}
  \nonumber [L_m,L_n]&=(m-n)L_{n+m}+{c_L\over 12}m(m^2-1)\delta_{m+n,0}\,,\\
  \nonumber [L_m,M_n]&=(m-n)M_{n+m}+{c_M\over 12}m(m^2-1)\delta_{m+n,0}\, ,\\ 
            [M_m,M_n]&=0\, ,    
  \end{align}
where the central charges $c_L$ and $c_M$ are given by the linear combination of the parent relativistic central charges
\begin{equation}\label{def of cC of CCFT}
c_L=\lim_{\epsilon\to 0 }(c^+-c^-)\, ,\qquad c_M=\lim_{\epsilon\to 0 }\epsilon(c^++c^-)\, .
\end{equation}
The algebra \eqref{BMS} which is given by the contraction of the Virasoro algebra in the boundary theory is exactly
the (centrally extended) BMS$_{3}$ algebra \cite{bar-comp}.

We would expect the same symmetry group for the CCFT dual to
the theory described by the action \eqref{Flataction} at the special point $m^2\ell^2=1/2$.
As we stated earlier, the $\epsilon\to 0$ limit in the boundary corresponds to the flat-space limit
or, more precisely, the $\kappa/\ell\to 0$ limit in the bulk side.
Using \eqref{central charge} and \eqref{def of cC of CCFT}, for the problem in hand we find
\be
 c_L=c_M=0\, .
 \ee

We will show that although the CCFT algebra has vanishing central charges,
it is possible to find a Cardy-like formula for the asymptotic growth of the number of states
which reproduces the entropy of the black hole \eqref{FlatBH}.
To add strength to this claim, let us find more evidence about
the correctness of our proposal using the CCFT energy-momentum tensor.

\subsubsection{Quasi local stress tensor}
The one-point function of the CCFT energy-momentum operator
corresponds to the quasilocal stress tensor of the bulk theory.
It was argued in \cite{Fareghbal:2013ifa} that the definition \eqref{def of BMS generators} provides a recipe
to calculate the components of the stress tensor in the asymptotically flat spacetimes.
Therefore, we can write
\begin{eqnarray}\label{def of energy-momentum }
 \nonumber \tilde T_{++}+\tilde T_{--}&=&\lim_{\epsilon\to 0}\epsilon\left(T_{++}+T_{--}\right)\, ,\\
  \nonumber\tilde T_{++}-\tilde T_{--}&=&\lim_{\epsilon\to 0}\left(T_{++}-T_{--}\right)\, ,\\
  \tilde T_{+-}&=&\lim_{\epsilon\to 0} T_{+-}\, ,
  \end{eqnarray}
where $T_{ij}$ and $\tilde T_{ij}$ are, respectively, the stress tensor of the asymptotically AdS and flat spacetimes
and $x^{\pm}$ are the light-cone coordinates constructed from the nonradial coordinates of the metrics.
In the above definition it was assumed that both the asymptotically AdS
and flat spacetimes are given in the BMS gauge \cite{Fareghbal:2013ifa}. 

The nonzero components of the stress tensor at the boundary
of the asymptotically AdS black hole \eqref{BHTbh} are given by \cite{Kwon:2011jz} 
\begin{align}
\nonumber T_{tt}&=\frac{1}{8\pi G \ell }\left(\frac{b^{2}\ell^{2}}{4}+4 M G\right), \\
\nonumber T_{t\phi}&=-\frac{a}{8\pi G \ell}\left(\frac{b^{2}\ell^{2}}{4}+4 M G\right), \\
          T_{\phi\phi}&=\frac{\ell}{8\pi G }\left(\frac{b^{2}\ell^{2}}{4}+4 M G\right).
\end{align}
The formula \eqref{def of energy-momentum } results in a stress tensor $\tilde T_{ij}$
for the asymptotically flat black hole \eqref{FlatBH} as follows:
\begin{equation}\label{flat stress tensor}
\tilde T_{tt}={b^2\over 32\pi}\, ,\qquad \tilde T_{\phi\phi}={\kappa^2 b^2\over 32 \pi}\, , \qquad \tilde T_{t\phi}=-{a b^2\over 32\pi}\, .
\end{equation}
Using $\tilde T_{ij}$ we can calculate the conserved charges of the black hole \eqref{FlatBH}.

 Let us denote the hypersurface of the spacetime where CCFT lives with $\partial {\cal M}$.
Its line element is given by taking the $\ell\to \infty$ limit of the conformal boundary \eqref{CB},
\be\label{bundary of flat}
ds_{\partial {\cal M}}^{2}=\frac{r^{2}}{\kappa^{2}}\left(-dt^{2}+\kappa^{2}d\phi^{2}\right).
\ee
Following Brown and York\rq{}s method \cite{Brown:1992br},
the charges associated to a boundary Killing vector $\xi^\mu$ are given by
\begin{equation}\label{BY formula}
Q_{\xi}=\int_{\Sigma} d\phi \sqrt{\sigma}\xi^\mu n^\nu \tilde T_{\mu\nu}\, ,
\end{equation}
where $\Sigma$ is the spacelike surface embedded in $\partial {\cal M}$ with induced metric $\sigma_{\mu\nu}$.
Moreover, $n^{\mu}$ is the timelike unit normal to $\Sigma$.
Using \eqref{bundary of flat} and \eqref{BY formula},
the mass and the angular momentum of the asymptotically flat black hole \eqref{FlatBH} are 
\begin{equation}\label{charges}
{\cal M}=Q_{\partial_t}={\kappa b^2\over 16}\, ,\qquad {\cal J}=Q_{\partial_\phi}=-{\kappa a b^2\over 16}\, .
\end{equation}
It is clear that $|{\cal J}|/{\cal M}=a$ as expected.

Given the expressions above, together with \eqref{Fcalentate},
it is straightforward to check that the first law of black hole thermodynamics is satisfied, i.e.
\be
d{\cal M}=T_{\rm flat}\,dS_{\rm flat}-\Omega_{{+}_{\rm flat}}\,d{\cal J}\, .
\ee

\subsubsection{Cardy-like formula}\label{entropy}
If the gravity theory \eqref{Flataction} has a dual description in terms of a CCFT,
then the entropy of the black hole \eqref{FlatBH} must be given by
the asymptotic growth of the number of states in the boundary theory.
In \cite{Bagchi:2012xr}, the authors found a Cardy-like formula
by computing the CCFT partition function using the saddle-point approximation.
However, in the recent papers \cite{Fareghbal:2014qga,Riegler:2014bia}
it was shown that the Cardy-like formula of \cite{Bagchi:2012xr} can be obtained
if one writes the Cardy formula in terms of CCFT parameters and then takes the $\epsilon\to 0$ limit.
In the current work we will use the same approach and take the $\epsilon\to 0$ limit from the Cardy formula in the parent CFT.
  
The CCFT algebra is given by \eqref{BMS}.
We denote the eigenvalues of $L_{0}$ and $M_{0}$ by $\Delta_L$ and $\Delta_M$, respectively.
For the current problem $c_{L}=c_{M}=0$, however, the eigenvalues of $L_{0}$ and $M_{0}$ are nonzero.
From the viewpoint of the limit \eqref{def of BMS generators} we see that the two labels $\Delta_L$ and $\Delta_M$
are related to the conformal weights in the two-dimensional CFT as
\footnote{
The Hilbert space construction of a CCFT is analogous to that of the relativistic 2D CFT.
Now, the states are labeled by the eigenvalues under $L_{0}$ and $M_{0}$.
We shall use the cylinder representation of the CCFT algebra \cite{Bagchi:2012cy}.
The Hilbert space of the 2D CCFT are constructed by considering the states having definite scaling dimensions.
We define primary states by demanding that the states in the theory be annihilated by all generators with $n>0$.
One can build up a tower of operators by acting on a primary state with the creation operators $L_{-n}$ and $M_{-n}$ ($n>0$).
}
\begin{align}\label{def of D_LM}
\Delta_{L}&=\lim_{\epsilon\to 0}\left(\Delta_{+}-\Delta_{-}\right)\, ,
\qquad \Delta_{M}=\lim_{\epsilon\to 0}\epsilon\left(\Delta_{+}+\Delta_{-}\right)\, .
\end{align}
We would like to remind the reader that the $\epsilon\to 0$ limit in the boundary field theory
corresponds to the $\kappa/\ell\to 0$ limit in the bulk.
Therefore, using \eqref{Delta+-} and \eqref{def of D_LM} one can easily find
\begin{equation}\label{CCFT quantities}
\Delta_{L}=\frac{a\kappa b^{2}}{16}\, ,\qquad \Delta_{M}=\frac{\kappa^2 b^{2}}{16}\, .
\end{equation}
Let us consider the Cardy formula \eqref{Cardy formula} and try to take its $\epsilon\to 0$  limit.
For our current problem the relativistic central charges are $c_+=c_-=3\epsilon$.
Using \eqref{CCFT quantities}, we obtain
\begin{align}\label{CCFT cardy}
\nonumber \lim_{\epsilon\to 0} {S_{\text{CFT}}}&= \lim_{\epsilon\to 0} 2\pi\left(\sqrt{c_+\Delta_+\over 6}+\sqrt{c_-\Delta_-\over 6}\right)\\
\nonumber &=\lim_{\epsilon\to 0} \pi\left[\sqrt{{\epsilon}\left({\Delta_M\over\epsilon}+\Delta_L\right)}+\sqrt{{\epsilon}\left({\Delta_M\over\epsilon}-\Delta_L\right)}\right]\\
&=2\pi\sqrt{\Delta_M}=S_{\text{CCFT}}\, .
\end{align}
This is  the Cardy-like formula for the CCFT dual to the flat-space limit of NMG.
Inserting (\ref{CCFT quantities}) into (\ref{CCFT cardy}), we finally recover the entropy (\ref{flat entropy}),
\be\label{Flat/CCFT}
S_{\text{CCFT}}=S_{\text{flat}}\, ,
\ee
as we wanted to show.
It is a quite nontrivial result since the theory has vanishing central charges.

\section{Conclusions}

In this paper, we have proposed a flat space generalization of the AdS$_3$/CFT$_{2}$ holographic correspondence.\footnote{We note that in \cite{Hasanpour:2011ji},
the authors considered an asymptotically flat geometry which is a solution to
three-dimensional Einstein gravity conformally coupled to a scalar field
and discussed gravity/CFT correspondence for this background.}
We have provided the first example of a holographic dual of an asymptotically flat black hole solution.
Due to the absence of black hole solutions in three-dimensional Einstein gravity with vanishing cosmological constant,
we have considered higher-derivative gravity theories which admit asymptotically flat black hole solutions.
The theory we have investigated is given by taking the flat-space limit ($\Lambda\to 0)$ of NMG.
We argued that the dual field theory of the black hole solution of this theory is a CCFT.
For this purpose, we have constructed a stress tensor for the
asymptotically flat black hole solution and computed the conserved charges.
We then verified it using the first principle of black hole thermodynamics.
Furthermore, we have used the Flat/CCFT correspondence to find the black hole entropy
in terms of the asymptotic growth of the number of CCFT states.

It is interesting to note that the symmetry algebra of the corresponding CCFT had vanishing central charges though
the asymptotic growth of  states were nonzero.
This remarkable point can be used for finding holographic duals of four-dimensional asymptotically flat spacetimes.
According to the proposal of Flat/CCFT correspondence, the dual of four-dimensional asymptotically flat black holes
are field theories with BMS$_{4}$ symmetry  \cite{Barnich:2009se, Barnich aspects}.
In \cite{Barnich:2011mi}, the authors constructed the field-dependent central extension of BMS$_{4}$ algebra
and found that for the Kerr black hole some of the charges involved divergent integrals on the 2-sphere
if they used extended BMS algebra with both supertranslations and superrotations.
Thus, at first sight, it seems that counting CCFT$_{3}$ states would be a problematic issue, but our current work
shows that counting the asymptotic growth of CCFT states can be done whatever the central charges are.

Although our current study gives a holographic description of asymptotically flat black holes in three-dimensional higher-derivative gravity,
we believe that the Flat/CCFT correspondence can be extended to find a holographic description of
black holes in higher dimensions and, specifically, the four-dimensional Kerr black hole.
We hope to explore other intriguing aspects of the relation between asymptotically flat spacetimes and CCFTs in our future works.

\section*{Acknowledgments}
We would like to thank Ali Naseh for useful discussions.
We are grateful to \'{A}lvaro V\'{e}liz-Osorio and Arjun Bagchi for their useful
comments on the revised version of the manuscript.
S.M.H. is supported in part by INFN.



\begin{thebibliography}{}
\bibitem{Bagchi:2010zz}
  A.~Bagchi,
  ``Correspondence between Asymptotically Flat Spacetimes and Nonrelativistic Conformal Field Theories,''
  Phys.\ Rev.\ Lett.\  {\bf 105}, 171601 (2010).

  A.~Bagchi,
  ``The BMS/GCA correspondence,''
  arXiv:1006.3354 [hep-th].

\bibitem{Bagchi:2012cy}
  A.~Bagchi and R.~Fareghbal,
  ``BMS/GCA Redux: Towards Flatspace Holography from Non-Relativistic Symmetries,''
  JHEP {\bf 1210}, 092 (2012)
  [arXiv:1203.5795 [hep-th]].

\bibitem{Krishnan:2013wta} 
  C.~Krishnan, A.~Raju and S.~Roy,
  ``A Grassmann path from $AdS_3$ to flat space,''
  JHEP {\bf 1403}, 036 (2014)
  [arXiv:1312.2941 [hep-th]].

\bibitem{Bagchi:2012xr} 
  A.~Bagchi, S.~Detournay, R.~Fareghbal and J.~Simon,
  ``Holography of 3d Flat Cosmological Horizons,''
  Phys.\ Rev.\ Lett.\  {\bf 110}, 141302 (2013)
  [arXiv:1208.4372 [hep-th]].

\bibitem{Cornalba:2002fi} 
  L.~Cornalba and M.~S.~Costa,
  ``A New cosmological scenario in string theory,''
  Phys.\ Rev.\ D {\bf 66}, 066001 (2002)
  [hep-th/0203031].

\bibitem{Detournay:2012ug} 
  S.~Detournay,
  ``Inner Mechanics of 3d Black Holes,''
  Phys.\ Rev.\ Lett.\  {\bf 109}, 031101 (2012)
  [arXiv:1204.6088 [hep-th]]. 
 
  A.~Castro and M.~J.~Rodriguez,
  ``Universal properties and the first law of black hole inner mechanics,''
  Phys.\ Rev.\ D {\bf 86}, 024008 (2012)
  [arXiv:1204.1284 [hep-th]].

\bibitem{Fareghbal:2014qga} 
  R.~Fareghbal and A.~Naseh,
  ``Aspects of Flat/CCFT Correspondence,''
  arXiv:1408.6932 [hep-th].

\bibitem{Riegler:2014bia} 
  M.~Riegler,
  ``Flat space limit of higher-spin Cardy formula,''
  Phys.\ Rev.\ D {\bf 91}, no. 2, 024044 (2015)
  [arXiv:1408.6931 [hep-th]].

\bibitem{Fareghbal:2013ifa} 
  R.~Fareghbal and A.~Naseh,
  ``Flat-Space Energy-Momentum Tensor from BMS/GCA Correspondence,''
  JHEP {\bf 1403}, 005 (2014)
  [arXiv:1312.2109 [hep-th]].

\bibitem{Czech:2012be}
  B.~Czech, J.~L.~Karczmarek, F.~Nogueira and M.~Van Raamsdonk,
  ``Rindler Quantum Gravity,''
  Class.\ Quant.\ Grav.\  {\bf 29}, 235025 (2012)
  [arXiv:1206.1323 [hep-th]].

\bibitem{Parikh:2012kg}
  M.~Parikh and P.~Samantray,
  ``Rindler-AdS/CFT,''
  arXiv:1211.7370 [hep-th].
  
\bibitem{Fareghbal:2014oba} 
  R.~Fareghbal and A.~Naseh,
  ``Rindler/Contracted-CFT Correspondence,''
  JHEP {\bf 1406}, 134 (2014)
  [arXiv:1404.3937 [hep-th]].

\bibitem{Bagchi:2014iea} 
  A.~Bagchi, R.~Basu, D.~Grumiller and M.~Riegler,
  ``Entanglement entropy in Galilean conformal field theories and flat holography,''
  Phys.\ Rev.\ Lett.\  {\bf 114}, no. 11, 111602 (2015)
  [arXiv:1410.4089 [hep-th]].

\bibitem{Ida:2000jh} 
  D.~Ida,
  ``No black hole theorem in three-dimensional gravity,''
  Phys.\ Rev.\ Lett.\  {\bf 85}, 3758 (2000)
  [gr-qc/0005129].

\bibitem{Wald:1993nt} 
  R.~M.~Wald,
  ``Black hole entropy is the Noether charge,''
  Phys.\ Rev.\ D {\bf 48}, 3427 (1993)
  [gr-qc/9307038].

\bibitem{Bergshoeff:2009hq} 
  E.~A.~Bergshoeff, O.~Hohm and P.~K.~Townsend,
  ``Massive Gravity in Three Dimensions,''
  Phys.\ Rev.\ Lett.\  {\bf 102}, 201301 (2009)
  [arXiv:0901.1766 [hep-th]].

\bibitem{Deser:2009hb} 
  S.~Deser,
  ``Ghost-free, finite, fourth order D=3 (alas) gravity,''
  Phys.\ Rev.\ Lett.\  {\bf 103}, 101302 (2009)
  [arXiv:0904.4473 [hep-th]].

\bibitem{Oliva:2009ip} 
  J.~Oliva, D.~Tempo and R.~Troncoso,
  ``Three-dimensional black holes, gravitational solitons, kinks and wormholes for BHT massive gravity,''
  JHEP {\bf 0907}, 011 (2009)
  [arXiv:0905.1545 [hep-th]].

\bibitem{Brown:1992br} 
  J.~D.~Brown and J.~W.~York, Jr.,
  ``Quasilocal energy and conserved charges derived from the gravitational action,''
  Phys.\ Rev.\ D {\bf 47}, 1407 (1993)
  [gr-qc/9209012].
 
\bibitem{Bergshoeff:2009aq} 
  E.~A.~Bergshoeff, O.~Hohm and P.~K.~Townsend,
  ``More on Massive 3D Gravity,''
  Phys.\ Rev.\ D {\bf 79}, 124042 (2009)
  [arXiv:0905.1259 [hep-th]].
  
\bibitem{Giribet:2009qz} 
  G.~Giribet, J.~Oliva, D.~Tempo and R.~Troncoso,
  ``Microscopic entropy of the three-dimensional rotating black hole of BHT massive gravity,''
  Phys.\ Rev.\ D {\bf 80}, 124046 (2009)
  [arXiv:0909.2564 [hep-th]].

\bibitem{deBuyl:2013ega} 
  S.~de Buyl, S.~Detournay, G.~Giribet and G.~S.~Ng,
  ``Baby de Sitter black holes and dS$_3$/CFT$_2$,''
  JHEP {\bf 1402}, 020 (2014)
  [arXiv:1308.5569 [hep-th]].

\bibitem{Brown:1986nw} 
  J.~D.~Brown and M.~Henneaux,
  ``Central Charges in the Canonical Realization of Asymptotic Symmetries: An Example from Three-Dimensional Gravity,''
  Commun.\ Math.\ Phys.\  {\bf 104}, 207 (1986).

\bibitem{BMS}
H.~Bondi, M.~G. van~der Burg, and A.~W. Metzner, ``Gravitational
waves in general relativity. 7. {W}aves from axisymmetric isolated systems,'' {\em
  Proc.\ Roy.\ Soc.\ Lond. A} {\bf 269} (1962)
21.

R.~K. Sachs, ``Gravitational waves in general relativity. 8.
{W}aves in asymptotically flat space-times,'' {\em Proc.\ Roy.\ Soc.\ Lond.\ A} {\bf
  270} (1962)
103.

R.~K. Sachs, ``Asymptotic symmetries in gravitational theory,''
{\em Phys.
  Rev.} {\bf 128} (1962) 2851.
  
\bibitem{Barnich aspects} 
G.~Barnich and C.~Troessaert,
  ``Aspects of the BMS/CFT correspondence,''
  JHEP {\bf 1005}, 062 (2010)
  [arXiv:1001.1541 [hep-th]].

\bibitem{bar-comp}
  G.~Barnich and G.~Compere,
  ``Classical central extension for asymptotic symmetries at null infinity in
  three spacetime dimensions,''
  Class.\ Quant.\ Grav.\  {\bf 24}, F15 (2007)
  [Erratum-ibid.\  {\bf 24}, 3139 (2007)]
  [arXiv:gr-qc/0610130].

\bibitem{Kwon:2011jz} 
  Y.~Kwon, S.~Nam, J.~D.~Park and S.~H.~Yi,
  ``Holographic Renormalization and Stress Tensors in New Massive Gravity,''
  JHEP {\bf 1111}, 029 (2011)
  [arXiv:1106.4609 [hep-th]].

\bibitem{Hasanpour:2011ji} 
  M.~Hasanpour, F.~Loran and H.~Razaghian,
  ``Gravity/CFT correspondence for three dimensional Einstein gravity with a conformal scalar field,''
  Nucl.\ Phys.\ B {\bf 867}, 483 (2013)
  [arXiv:1104.5142 [hep-th]].

\bibitem{Barnich:2009se} 
  G.~Barnich and C.~Troessaert,
  ``Symmetries of asymptotically flat 4 dimensional spacetimes at null infinity revisited,''
  Phys.\ Rev.\ Lett.\  {\bf 105}, 111103 (2010)
  [arXiv:0909.2617 [gr-qc]].

\bibitem{Barnich:2011mi} 
  G.~Barnich and C.~Troessaert,
  ``BMS charge algebra,''
  JHEP {\bf 1112}, 105 (2011)
  [arXiv:1106.0213 [hep-th]].





\end{thebibliography}
\end{document}